\documentclass[preprint,showpacs,preprintnumbers,amsmath,amssymb]{revtex4}
\usepackage{graphicx,epsfig,dcolumn,bm,epic,eepic,float}
\usepackage{amsmath}
\usepackage{latexsym}
\usepackage{color}
\usepackage{makeidx,shortvrb,latexsym}
\begin{document}
\unitlength 1 cm
\newcommand{\be}{\begin{equation}}
\newcommand{\ee}{\end{equation}}
\newcommand{\bearr}{\begin{eqnarray}}
\newcommand{\eearr}{\end{eqnarray}}
\newcommand{\nn}{\nonumber}
\newcommand{\vk}{\vec k}
\newcommand{\vp}{\vec p}
\newcommand{\vq}{\vec q}
\newcommand{\vkp}{\vec {k'}}
\newcommand{\vpp}{\vec {p'}}
\newcommand{\vqp}{\vec {q'}}
\newcommand{\bk}{{\bf k}}
\newcommand{\bp}{{\bf p}}
\newcommand{\bq}{{\bf q}}
\newcommand{\br}{{\bf r}}
\newcommand{\bR}{{\bf R}}
\newcommand{\up}{\uparrow}
\newcommand{\down}{\downarrow}
\newcommand{\fns}{\footnotesize}
\newcommand{\ns}{\normalsize}
\newcommand{\cdag}{c^{\dagger}}
\title {A phenomenological investigation of the beauty content of a proton in the framework of $k_t$-factorization using $KMR$ and $MRW$ unintegrated parton distributions }
\author{N. Olanj}\altaffiliation {Corresponding author, Email: {$n\_olanj@basu.ac.ir$}, Tel:+98-81-31400000}
\affiliation{Department of Physics, Faculty of Science, Bu-Ali
Sina University, 65178, Hamedan, Iran} 
\begin{abstract}
In this paper, we address the reduced beauty cross section ($\sigma_{red}^{b\overline{b}}(x, Q^2)$) and the beauty structure function ($F_2^{b\overline{b}}(x, Q^2)$), to study the beauty content of a proton. We calculate $\sigma_{red}^{b\overline{b}}$ and $F_2^{b\overline{b}}$ in the $k_t$-factorization formalism by using the integral form of the $Kimber$-$Martin$-$Ryskin$ and $Martin$-$Ryskin$-$Watt$ unintegrated parton distribution function ($KMR$ and $MRW$-$UPDF$) with the angular ordering constraint ($AOC$) and the $MMHT2014$ $PDF$ set as the input. Recently Guiot and van Hameren demonstrated that the upper limit, $k_{max}$, of the transverse-momentum integration performed in the $k_t$-factorization formalism should be almost equal to $Q$, where $Q$ is the hard scale, otherwise it leads to an overestimation of the proton structure function ($F_2(x, Q^2)$).
In the present work, we show that $k_{max}$ cannot be equal to $Q$ at low and moderate energy region, and also by considering the gluon and quark contributions to the same perturbative order and a physical gauge for the gluon,
i.e., $A^{\mu}q^{\prime}_{\mu}=0 $ in the calculation of $F_2^{b\overline{b}}$ in the $k_t$-factorization formalism, we do not encounter any overestimation of the theoretical predictions due to different choices of $k_{max}>Q$.
Finally, the resulted $\sigma_{red}^{b\overline{b}}$ and $F_2^{b\overline{b}}$ are
compared to the experimental data and the theoretical
predictions. In general, the extracted
$\sigma_{red}^{b\overline{b}}$ and $F_2^{b\overline{b}}$ based on the $KMR$ and $MRW$ approaches are in perfect agreement with the experimental data and theoretical
predictions at high energies, but at low and moderate energies, the one developed
from the $KMR$ approach has better consistency than that of $MRW$ approach.
\end{abstract}
 \maketitle
\section{Introduction}

The study of the charm and beauty content of a proton in deep inelastic $ep$ scattering at $HERA$ plays an important role in the investigation of  the theory of the perturbative quantum $chromodynamics$ ($pQCD$) at the small $Bjorken$ scale ($x$) \cite{3p, 4, pair}, the electroweak Higgs boson production at the $LHC$ \cite{2} and hadron-hadron differential cross sections.

Recently,  we investigated the charm content of a proton in the frameworks of the
$KMR$ \cite{9} and $MRW$ \cite{10} approaches by calculating the charm structure function ${F_2}^{c\overline{c}}(x,Q^2)$ \cite{charm}
 in the $k_t$-factorization formalism \cite{new1, new2, new3, new4, new5}. We showed that the calculated charm structure functions by 
 using the $MRW$-$UPDF$ and $KMR$-$UPDF$ are consistent with the experimental data and the theoretical predictions based on the
  general-mass variable-flavor-number scheme ($GMVFNS$) \cite{29}, the $LO$ collinear procedure and the saturation model introduced
   by $Golec-Biernat$ and $W\ddot{u}sthoff$ ($GBW$) \cite{GBW}. Also, in the reference \cite{Lip}, the $b$-quark contribution to the inclusive proton structure function $F_2(x, Q^2 )$ at high values of $Q^2$ has been investigated at the leading-order $k_t$-factorization approach using $KMR$-$UPDF$ and only considering the gluon contribution.

 Recently, in the reference \cite{hera2018}, measurements of charm and beauty production cross section in deep inelastic $ep$ scattering at
$HERA$ from the $H1$ and $ZEUS$ Collaborations are combined and results for the so-called 
reduced charm and beauty cross section ($\sigma_{red}^{q\overline{q}}(x, Q^2), q= c, b$) are obtained in the kinematic range of negative 
four-momentum transfer squared ($-q^2=Q^2$) of the photon $2.5 GeV^2 \leq Q^2 \leq 2000 GeV^2$ and Bjorken 
scaling variable $3\times10^{-5} \leq x_{Bj }\leq 5 \times 10^{-2}$. The double-differential cross section for the production of a heavy
flavour of type $q$ ($ q= c, b$) may then be written in terms of  the heavy-flavour contributions of the structure functions $F_2(x, Q^2)$ ($F_T(x, Q^2)+F_L(x, Q^2)$) 
and $F_L(x, Q^2)$ \cite{tkimber, 28tkimber}, as follows:
 \begin{eqnarray}
\frac{d^2\sigma^{q\overline{q}}}{dx dQ^2}&=&\frac{4\pi\alpha^2(Q^2)}{Q^4}\Bigg[\frac{1+(1-y)^2}{2x}F_T^{q\overline{q}}(x, Q^2)+\frac{1-y}{x}F_L^{q\overline{q}}(x, Q^2)\Bigg]\nonumber\\&=&\frac{2\pi\alpha^2(Q^2)}{xQ^4}\Bigg[(1+(1-y)^2)F_2^{q\overline{q}}(x, Q^2)-y^2F_L^{q\overline{q}}(x, Q^2)\Bigg],
 \label{eq:1n}
\end{eqnarray}
where $y=\frac{Q^2}{xs}$ ($s$ is $CM$ energy squared) denotes the lepton inelasticity, the fraction of energy transferred from the electron
 in the fixed proton frame. The reduced cross sections are defined, as follows:
 \begin{eqnarray}
\sigma_{red}^{q\overline{q}}(x, Q^2)&=&\frac{d^2\sigma^{q\overline{q}}}{dx dQ^2}\cdot \frac{xQ^4}{2\pi \alpha^2(Q^2)(1+(1-y)^2)}\nonumber\\&=&F_2^{q\overline{q}}(x, Q^2)-\frac{y^2}{1+(1-y)^2} F_L^{q\overline{q}}(x, Q^2).
 \label{eq:2n}
\end{eqnarray}

In expressing the importance of the investigation of the proton beauty contents in this paper, it should be noted that the reduced cross section ($\sigma_{red}^{q\overline{q}}(x, Q^2)$) is dependent on the heavy-flavour longitudinal structure
 function ($F_L^{q\overline{q}}(x, Q^2)$). Therefore, since the longitudinal structure function is directly sensitive to the
  gluon distributions, the calculations of the reduced cross section are beyond the standard collinear factorization procedure
   i.e. the $k_t$-factorization formalism.

In this work, we use the integral form of $MRW$-$UPDF$ and $KMR$-$UPDF$  with the angular ordering constraint ($AOC$) \cite{221} and the ordinary parton distribution functions (the cutoff independent $PDF$) according to the investigations carried out in the references \cite{221} and \cite{222} as input in the $k_t$-factorization formalism to calculate the reduced cross section of production of a beauty quark pair in the final state of the deep inelastic $ep$ scattering, ($\sigma_{red}^{b\overline{b}}(x, Q^2)$) and the beauty structure function ($F_2^{b\overline{b}}(x, Q^2)$).
It is worth mentioning that in the references \cite{221} and \cite{222}, it is stated that " the differential version of $KMR$ prescription and the implementations of angular (strong) ordering constraints ($AOC$ ($SOC$)), cause the negative-discontinuous $UPDF$ with the ordinary parton distribution functions as the input and finally leads to results far from experimental data, but those the proton (longitudinal) structure functions calculated based on the integral prescription of the $KMR$-$UPDF$ with the $AOC$ and the ordinary $PDF$ as the input are reasonably consistent with the experimental data."
Then the predictions of these two approaches by using the $MMHT2014$-$LO$  and $MMHT2014$-$NLO$ set of the $PDF$ \cite{22} as
 input for the reduced beauty cross section are compared to the combined data of the $H1$ and $ZEUS$ Collaborations at
$HERA$ \cite{hera2018} and theoretical predictions based on
the $HERAPDF2.0$ $FF3A$ set \cite{FF3A}. Also, the resulted beauty structure function is compared to the
predictions of the $ MSTW08-NLO$ $QCD$ fits \cite{z1} and the $ZEUS$ measurements \cite{z2,z3,z4}. As shown in references \cite{16, badieian}, $UPDF$ with different input $PDF$ sets are almost
 very similar and stable, so we can still use $MMHT2014$ $PDF$ set instead of the new $MSHT20$ $PDF$ set \cite{MSHT20} in this work. In general,
  it is shown that the calculated reduced beauty cross section ( $\sigma_{red}^{b\overline{b}}$)  and the beauty structure function ($F^{b\overline{b}}$) based on the $UPDF$ of the two approaches are very consistent
   with the experimental data, especially at high energies. However, the reduced beauty cross sections and the beauty structure functions, which are extracted from the $KMR$ approach, have a better agreement
   with the experimental data with respect to that of $MRW$ at low and moderate energies.

It should be noted that the $k_t$-factorization formalism is computationally more straightforward than the theory of the $pQCD$. 
 The discrepancy between the $pQCD$ and the $k_t$-factorization prediction can be reduced by refitting the input integrated 
 $PDF$ \cite{WattWZ} and using the cut-off dependent $PDF$ \cite{Lotfi}. As explained in the reference \cite{WattWZ},
  this treatment is adequate for initial investigations and descriptions of exclusive processes.

It is worth mentioning that recently Guiot and van Hameren ($GvH$) encountered an overestimation of the exact structure function by 
calculate the proton structure function ($F_2(x, Q^2)$) in the $k_t$-factorization formalism at order $O(\lambda^2)$, with $\lambda$ the coupling of the Yukawa theory by using the differential form of the $KMRW-UPDF$ computed in the Yukawa theory only considering the quark contributions \cite{GvH}. Therefore, $GvH$ argued  that the upper limit, $k_{max}$, of the transverse-momentum integration performed in the $k_t$-factorization formalism is equal to $\mu_F\sim Q$ ($Q$ is the hard scale) used to factorize the cross section into an off-shell hard coefficient and a universal factor.
In the present work, we show that $k_{max}$ cannot be equal to $Q$ at low and moderate energy region ($2.5 GeV^2 \leq Q^2 \leq 120 GeV^2$), and also by considering the gluon and quark contributions to the same perturbative order and a physical gauge for the gluon,
i.e., $A^{\mu}q^{\prime}_{\mu}=0 $ in the calculation of $F_2^{b\overline{b}}(x, Q^2)$ and $F_L^{b\overline{b}}(x, Q^2)$  in the $k_t$-factorization formalism, we do not encounter any overestimation of the theoretical predictions due to different choices of $k_{max}>Q$.

Due to the importance of this subject, in our previous articles, we  investigated the general behavior and
stability of  the $KMR$ and $MRW$ approaches \cite{11,12,13,14,15,16,17,18,19} and in this paper, we  study the beauty content of 
a proton by examining the reduced cross section ($\sigma_{red}^{b\overline{b}}(x, Q^2)$ ) and the beauty structure function ($F^{b\overline{b}}(x, Q^2)$). Also, we have successfully used $KMR$-$UPDF$ in our previous articles,
 to calculate the inclusive production of the $W$ and $Z$ gauge vector bosons  \cite{171,181}, the semi-$NLO$ production of $Higgs$
  bosons  \cite{191}, the production of forward-center and forward-forward di-jets \cite{201}, the prompt-photon
   pair production \cite{211}, the single-photon production \cite{amin} and  the charm structure function\cite{charm}. 
   We explored the phenomenology of  the integral and the differential versions of the $KMR$-$UPDF$ using the angular (strong) 
   ordering ($AOC$ ($SOC$)) constraints in the reference \cite{221}. Also, among the applications of these $UPDF$, one can refer to the references \cite{s1,s2,s3,s4}.

So, the paper is organized as follows:  an overview of the $KMR$ and $MRW$ approaches to generating $UPDF$ and calculation
 of the beauty contribution to the proton structure function
 ($F_{2}^{b\overline{b}}(x, Q^2)$) and the proton longitudinal structure function ($F_{L}^{b\overline{b}}(x, Q^2)$) based on the $k_t$-factorization formalism are provided in section $II$. Finally, the results of the reduced beauty cross section and the beauty structure function 
  in the $k_t$-factorization formalism using the $KMR$-$UPDF$ and $MRW$-$UPDF$ as input are presented in section $III$.
\section{$KMR$-$UPDF$, $MRW$-$UPDF$ approaches and  $F_{2}^{b\overline{b}}(x, Q^2)$ and $F_{L}^{b\overline{b}}(x, Q^2)$
 in the $k_t $-factorization formalism}   
A brief review of the $KMR$ \cite{9} and $MRW$ \cite{10} approaches to generating $UPDF$ ($f_{a}(x, k_{t}^2, \mu^{2})$ at the $LO$ and   $NLO$ levels, respectively, where
$x$, $k_t$, and $\mu$ are the longitudinal momentum fraction, the transverse momentum, and the factorization
scale, respectively) is provided in this
section. The $KMR$ and $MRW$ formalisms are based on the $DGLAP$ equations using some modifications
due to the separation of the virtual and real parts of the evolutions.

The $KMR$ approach leads to the following integral forms for the quark and gluon $UPDF$ at the $LO$ level, respectively:

\begin{eqnarray}
f_{q}(x,k_{t}^2,\mu^{2})&=&T_q(k_t,\mu)\frac{\alpha_s({k_t}^2)}{2\pi}
\nonumber\\&\times&
\int_x^{1-\Delta}dz\Bigg[P_{qq}(z)\frac{x}{z}\,q\left(\frac{x}{z} ,
{k_t}^2 \right)\cr &+& P_{qg}(z)\frac{x}{z}\,g\left(\frac{x}{z} ,
{k_t}^2 \right)\Bigg],
 \label{eq:8}
\end{eqnarray}
\begin{eqnarray}
f_{g}(x,k_{t}^2,\mu^{2})&=&T_g(k_t,\mu)\frac{\alpha_s({k_t}^2)}{2\pi}
\nonumber\\&\times& \int_x^{1-\Delta}dz\Bigg[\sum_q
P_{gq}(z)\frac{x}{z}\,q\left(\frac{x}{z} , {k_t}^2 \right) \cr &+&
P_{gg}(z)\frac{x}{z}\,g\left(\frac{x}{z} , {k_t}^2 \right)\Bigg],
 \label{eq:9}
\end{eqnarray}
 where $P_{aa^{\prime}}(x)$ are the corresponding
splitting functions and the survival probability factors, $T_a$,
is evaluated from:
\begin{eqnarray}
T_a(k_t,\mu)&=&\exp\Bigg[-\int_{k_t^2}^{\mu^2}\frac{\alpha_s({k'_t}^2)}{2\pi}\frac{{dk'_t}^{2}}{{k'_t}^{2}}
\cr &\times& \sum_{a'}\int_0^{1-\Delta}dz'P_{a'a}(z')\Bigg],
 \label{eq:5}
\end{eqnarray}
where $\Delta$ is a cutoff to prevent the integrals from becoming singular at $z = 1$ (arises from the soft gluon emission).  By considering
 the angular ordering constraint ($AOC$), which is the consequence of the coherent gluon emissions, the cutoff is equal to
  $\frac{k_t}{\mu+k_t}$ and $UPDF$ extend smoothly into the domain $k_t>\mu$. It should be mentioned that in this approach, $T_a$ is
considered to be unity for $k_t>\mu$. Therefore $k_{max}=\mu \sim Q$ is not an intrinsic property of the the unintegrated parton distribution function.

The $MRW$ approach leads to the following integral forms for the quark and gluon $UPDF$ at the $NLO$ level:
\begin{eqnarray}
f_{a}(x,k_{t}^2,\mu^{2})&=&\int_x^{1}dz
T_a(k^{2},\mu^{2})\frac{\alpha_s({k}^2)}{2\pi} \nonumber\\&\times&
\sum_{b=q,g}P_{ab}^{(0+1)}(z)\,b\left(\frac{x}{z} , {k}^2
\right)\Theta(\mu^{2}-k^{2}),
 \label{eq:10}
\end{eqnarray}
where
\begin{eqnarray}
P_{ab}^{(0+1)}(z)&=&P_{ab}^{(0)}(z)+\frac{\alpha_s}{2\pi}P_{ab}^{(1)}(z),\nonumber\\k^2&=&\frac{k_t^2}{1-z}
 \label{eq:11},
\end{eqnarray}
and
\begin{eqnarray}
T_a(k^{2},\mu^{2})&=&\exp\Bigg(-\int_{k^2}^{\mu^2}\frac{\alpha_s({\kappa}^2)}{2\pi}\frac{{d\kappa}^{2}}{{\kappa}^{2}}
\cr &\times& \sum_{b=q,g}\int_0^{1}d\zeta \zeta
P_{ba}^{(0+1)}(\zeta)\Bigg).
 \label{eq:12}
\end{eqnarray}
 $P_{ab}^{(0)}$ and $P_{ab}^{(1)}$ functions in the above
 equations
correspond to the $LO$ and $NLO$ contributions of the splitting
functions, respectively, which are given in the reference \cite{25}.  In the $MRW$ approach, unlike the $KMR$ approach, the cuttoff is
 imposed only on the terms in which the splitting functions
are singular, i.e., the terms that include $P_{qq}$ and $P_{gg}$, also, the scale $k^2=\frac{k_t^2}{1-z}$ is used instead of the scale $k_t^2$. For more details see reference \cite{14}.

In the following, we briefly present the formulations of the beauty structure function
 ($F_{2}^{b\overline{b}}(x, Q^2)$) and the beauty longitudinal structure function ($F_{L}^{b\overline{b}}(x, Q^2)$) in
 the $k_t $-factorization formalism. By considering the gluon and quark contributions to the same perturbative order and a physical gauge for the gluon,
i.e., $A^{\mu}q^{\prime}_{\mu}=0 $ $(q^\prime=q+xp)$, the beauty structure function $F_{2}^{b\overline{b}}(x, Q^2)$ is given by the sum of
the gluon contribution (the subprocess $g \rightarrow q \overline{q}$, the equation (\ref{eq:2})) and the quark contribution
 (the subprocess $q \rightarrow qg$, the equation
(\ref{eq:d})) according to the equations (8) and (12) of the reference \cite{charm}.

For the gluon contribution:
\begin{eqnarray}
{F_2}^{b\overline{b}}_{g \rightarrow q \overline{q}}(x,Q^2) &=&
e_b^2 \frac{Q^2}{4\pi} \int_{k_0^2}^{k_{max}^2}\frac{dk_t^2}{k_t^4}\int_0^{1}d\beta\int_{k_0^2}^{k_{max}^2}
d^2\kappa_t \alpha_s(\mu^2) f_g\left(\frac{x}{z},k_t^2,\mu^2\right)
 \Theta(1-\frac{x}{z})\nonumber\\
\Bigg \lbrace  [\beta^2 &+& (1-\beta^2)] (
\frac{\bf{\kappa_t}}{D_1}-\frac{(\bf{\kappa_t}-\bf{k_t})}{D_2} )^2 +
[m_b^2+4Q^2\beta^2(1-\beta)^2] (\frac{1}{D_1}-\frac{1}{D_2})^2 \Bigg
\rbrace
 \label{eq:2},
\end{eqnarray}
where
\begin{eqnarray}
D_1&=&\kappa_t^2+\beta(1-\beta)Q^2+m_b^2,\nonumber\\D_2&=&({\bf{\kappa_t}}
-{\bf{k_t}})^2 +\beta(1-\beta)Q^2+m_b^2
 \label{eq:3},
\end{eqnarray}
and
\begin{eqnarray}
\frac{1}{z}=1+\frac{\kappa_{t}^{2}+m_b^2}{(1-\beta)Q^2}+\frac{k_t^2+\kappa_t^2-2{\bf{\kappa_t}}.{\bf{k_t}}+m_b^2}{\beta
Q^2}
 \label{eq:a},
\end{eqnarray}
where in the above equations, the variable $\beta$ is defined as the light-cone fraction of the photon momentum carried
by the internal quark and $k_0$ is chosen to be about 1 $GeV$. The graphical representations of $k_t$ and $\kappa_t$ are introduced in the figure 7 of the reference \cite{14}. The scale $\mu$ controls both the
unintegrated partons and the $QCD$ coupling constant ($\alpha_s$)
and it is chosen as follows:
\begin{eqnarray}
\mu^2=k_t^2+\kappa_t^2+m_b^2
 \label{eq:b}.
\end{eqnarray}

It should be mentioned that the imposition of angular ordering constraint ($AOC$) at the last step of the evolution instead of the strong ordering constraint ($SOC$) leads to physically reasonable unintegrated parton distribution functions which extend smoothly into the domain $k_t>\mu$ \cite{9}. Therefore, the acceptable value of $k_{max}$ is the value that does not change the result of structure function by increasing it. For example, in the reference \cite{tkimber}, $k_{max}$ is considered equal to $4Q$.

For the quark contribution:
\begin{eqnarray}
{F_2}^{b\overline{b}}_{q \rightarrow qg}(x,Q^2)= &e_b^2&
\int_{k_0^2}^{Q^2}
\frac{d\kappa_t^2}{\kappa_t^2}\frac{\alpha_s(\kappa_t^2)}{2\pi}\int_{k_0^2}^{\kappa_t^2}\frac{dk_t^2}{k_t^2}
\int_{x}^{\frac{Q}{(Q+k_t)}}dz\nonumber\\ &\Bigg[& f_b
\left(\frac{x}{z} , k_t^2,Q^2\right)+f_{\overline{b}}
\left(\frac{x}{z} , k_t^2,Q^2\right) \Bigg] P_{qq}(z)
 \label{eq:d}.
\end{eqnarray}
In this paper, the mass of beauty quark is considered to be $m_b = 4.18GeV$. See reference \cite{charm} for more details.

As mentioned in reference \cite{charm}, the dominant mechanism of the proton $c,b$-quark electroproduction is the subprocess
 $g\rightarrow qq$, and since we are working in the small $x$ region (i.e. the high energy region), we ignored the contribution 
 of the non-perturbative region. According to the above, the beauty longitudinal structure function ($F_{L}^{b\overline{b}}(x, Q^2)$) in
 the $k_t $-factorization approach is presented as follows:
 \begin{eqnarray}
F_{L}^{b\overline{b}} (x, Q^2) =&\frac{Q^4}{\pi^2}& e_b^2
 \int_{k_0^2}^{k_{max}^2}\frac{dk_t^2}{k_t^4} \Theta(k^2-k_0^2)\int_0^{1}d\beta\int_{k_0^2}^{k_{max}^2}
d^2\kappa_ t \alpha_s(\mu^2) \beta^2(1-\beta)^2\left
(\frac{1}{D_1}-\frac{1}{D_2}\right)^2\nonumber\\ &\times &
f_g\left(\frac{x}{z}, k_t^2, \mu^2\right) + e_b^2 \frac{\alpha_s(Q^2)}{\pi} \frac{4}{3}
\int_x^{1}\frac{dy}{y}(\frac{x}{y})^2
[q(y, Q^2)+\overline{q}(y, Q^2)]
\label{eq:olanj},
\end{eqnarray}
where $y=x\bigg(1+\frac{{\kappa_t^{'}}^2+m_b^2}{\beta(1-\beta)Q^2}\bigg)$ 
(in which $\bf \kappa_t^{'}=\bf \kappa_t- (1-\beta)\bf k_t$) and the variables of the above equation are the same as the variables of
 the beauty structure function $(F_{2}^{b\overline{b}}(x, Q^2))$. It should be noted that the first term is derived with
the use of a pure gluon contribution from the perturbative region in the $k_t $-factorization approach. The second term is the beauty quark contribution in the longitudinal structure function which comes from the collinear factorization.

\section{Results, discussions and conclusions}
As mentioned before, the purpose of this work is a detailed investigation of the beauty content of a proton in
 the framework of $k_t$-factorization using $KMR$ and $MRW$ approaches to generate the $UPDF$, validate these two approaches and also investigatigation the upper limit of transverse momentum ($k_{max}$) in the $k_t $-factorization formalism. For this purpose,
the reduced beauty cross sections ($\sigma_{red}^{b\overline{b}}(x, Q^2)$, the equation (\ref{eq:2n})) are calculated
   by using the beauty structure functions
 ($F_{2}^{b\overline{b}}(x, Q^2)$, the sum of the equations (\ref{eq:2}) and (\ref{eq:d})) and the beauty longitudinal 
 structure functions ($F_{L}^{b\overline{b}}(x, Q^2)$, the equation (\ref{eq:olanj})) in
 the $k_t $-factorization formalism. The  integral form of $KMR$-$UPDF$ and $MRW$-$UPDF$, i.e.,
  the equations (\ref{eq:8}), (\ref{eq:9}) and  (\ref{eq:10}) with the angular ordering constraint ($AOC$) are used as input of the beauty structure functions and
   the beauty longitudinal structure functions in the $k_t $-factorization formalism with different $k_{max}\geq Q$.

In the figures 1, the reduced beauty cross section ($\sigma_{red}^{b\overline{b}}$) are displayed  in the framework of $k_t$-factorization using the $KMR$ and $MRW$ approaches as a function of $x$ for different values of $Q^2$= $2.5, 5, 7, 12, 18, 32, 60, 120, 200, 350, 650$ and $2000$ $GeV^2$  with the input
  $MMHT2014$ set of $PDF$ (to generate the $UPDF$) at the $LO$ and
  $NLO$ approximations, respectively, with $k_{max}^2=Q^2, 16Q^2$ in all panels, $k_{max}^2=36Q^2$ in panels a, b and c and $k_{max}^2=10^4 GeV^2$  in panels c and i. These results are compared to the combined data of the $H1$ and $ZEUS$ Collaborations at
$HERA$ \cite{hera2018} (the full circle points) and theoretical predictions based on
the $HERAPDF2.0$ $FF3A$ set \cite{FF3A} (dash-dot curves).
In the figure 2, the obtained results from the calculations of the beauty structure functions are presented as a function of $Q^2$ for various $x$ values using the $KMR$ ( $MMHT2014-LO$ $PDF$, dash curves) and $MRW$ ( $MMHT2014-NLO$ $PDF$, full curves) approaches with $k_{max}^2=Q^2$ ( at $i=0, 2$ and $7$), $16Q^2$ ( at all $i$) and $10^4 GeV^2$ ( at $i=2$ and $7$).
These results are compared to the experimental measurements of $ZEUS$ (filled circles \cite{4}, open circles \cite{z3} and open triangles \cite{z4}) and the predictions of $MSTW08-NLO$ $QCD$ calculations \cite{z1}.
To provide a clear comparison between the frameworks of the $KMR$ and $MRW$ approaches, we have plotted the $KMR-UPDF$ (dash curves) and $MRW-UPDF$ (full curves) versus ${k_t}^2$ at typical values of $x = 0.01, 0.001$ and $0.0001$ and the factorization scales $Q^2=60$  and $350$ $GeV^2$ for the beauty and gluon partons, in the figure 3.  Also, the beauty and gluon $PDF$ at scales $Q^2=60$ and $350$ $GeV^2$ are plotted by using the $MMHT2014$-$LO$ (dash curves) and $MMHT2014$-$NLO$ (full curves) \cite{22} in figure 4.
It should be mentioned that in the calculations related to the figures 1, 2 and 3, we consider the $QCD$ coupling constant,  
$\alpha_s({M_z}^2)$, to be the same as those used in fitting the input $PDF$ to the 
$KMR-UPDF$ and $MRW-UPDF$, i.e. 
$\alpha_{s, LO}({M_z}^2)=0.135$ and $\alpha_{s, NLO}({M_z}^2)=0.118$, respectively.

In general, the extracted
$\sigma_{red}^{q\overline{q}}$ and $F_{2}^{b\overline{b}}(x, Q^2)$ based on both the $KMR$ and $MRW$ approaches are in perfect consistent with the experimental data \cite{hera2018, z1} and the theoretical
predictions \cite{4, FF3A, z3, z4} at high energies, but the one developed
from the $KMR$ approach has a better agreement with
the experimental data and the theoretical
predictions with respect to that of $MRW$ approach at low and moderate energies.

As shown in the figures 1 and 2, and we expected according to the figures 3 and 4 ( see panels d and j of the figure 3 (the large $x$ and high energy region) and the figure 4), the results of the $KMR$ and $MRW$ approaches are very close to each other at the high hard scale ($Q^2$) and large $x$, but they become separated as the hard scale and $x$ decrease. It should be noted that this decrease in difference with increasing hard scale $Q$ and $x$ is due to the use of the scale $k^2=\frac{k_t^2}{1-z}$, the coupling constant $\alpha_{s,NLO}({M_z}^2)=0.118$ and $MMHT2014$-$NLO$ $PDF$ set instead of the scale $k_t^2$, the coupling constant $\alpha_{s,LO}({M_z}^2)=0.135$ and $MMHT2014$-$LO$ $PDF$ set in the $MRW$ approach. Note that this decrease cannot be a result of different use of cut-off and splitting functions.

It is clear in the figures 1 and 2 that at low and moderate $x$ and low energy region (see the first 8 panels of the figure 1, $Q^2\leq 120 GeV^2$and $i=7$ in the figure 2), there is no good agreement between the experimental data and the obtained results considering $k_{max}^2=Q^2$ and this is due to the non-negligibility value of $KMR$-$UPDF$  in the $k_t >Q$ region at low $x$ and $Q^2$ ( see panels c and i of the figure 3). But the results obtained from both $KMR$ and $MRW$ approaches, considering $K_{max}^2\geq 16Q^2$, have a good agreement with the theoretical predictions in all panels of the figures 1 and 2 (as mentioned in reference \cite{tkimber}).
 Also, at large $x$ and high energy region (see the last 4 panels of the figure 1, $Q^2\geq 200 GeV^2$ and $i=2$ in the figure 2), with the increase of $k_{max}^2$ from $Q^2$ to $10^4$, the results are almost the same and this is due to the negligible value of $KMR$-$UPDF$ and $MRW$-$UPDF$ in the $k_t >Q$ region at large $x$ and high values of $Q^2$ ( see panels d and j of the figure 3). It should be noted that according to the figures 1 and 2, we do not encounter any overestimation of the theoretical predictions with increasing $k_{max}$.

Also, as it is clear in the figures 1 and 2, at high (low) energy region, the result of the reduced beauty cross section and the beauty structure functions calculations have not changed by increasing $k_{max}$ from $Q$ to $10Q$ (from $4Q$ to $10Q$), so $k_{max}$=$Q$ ($k_{max}$=$4Q$) can be considered to save calculation time at high (low) energy.
\begin{figure}[ht]
\includegraphics[width=160mm]{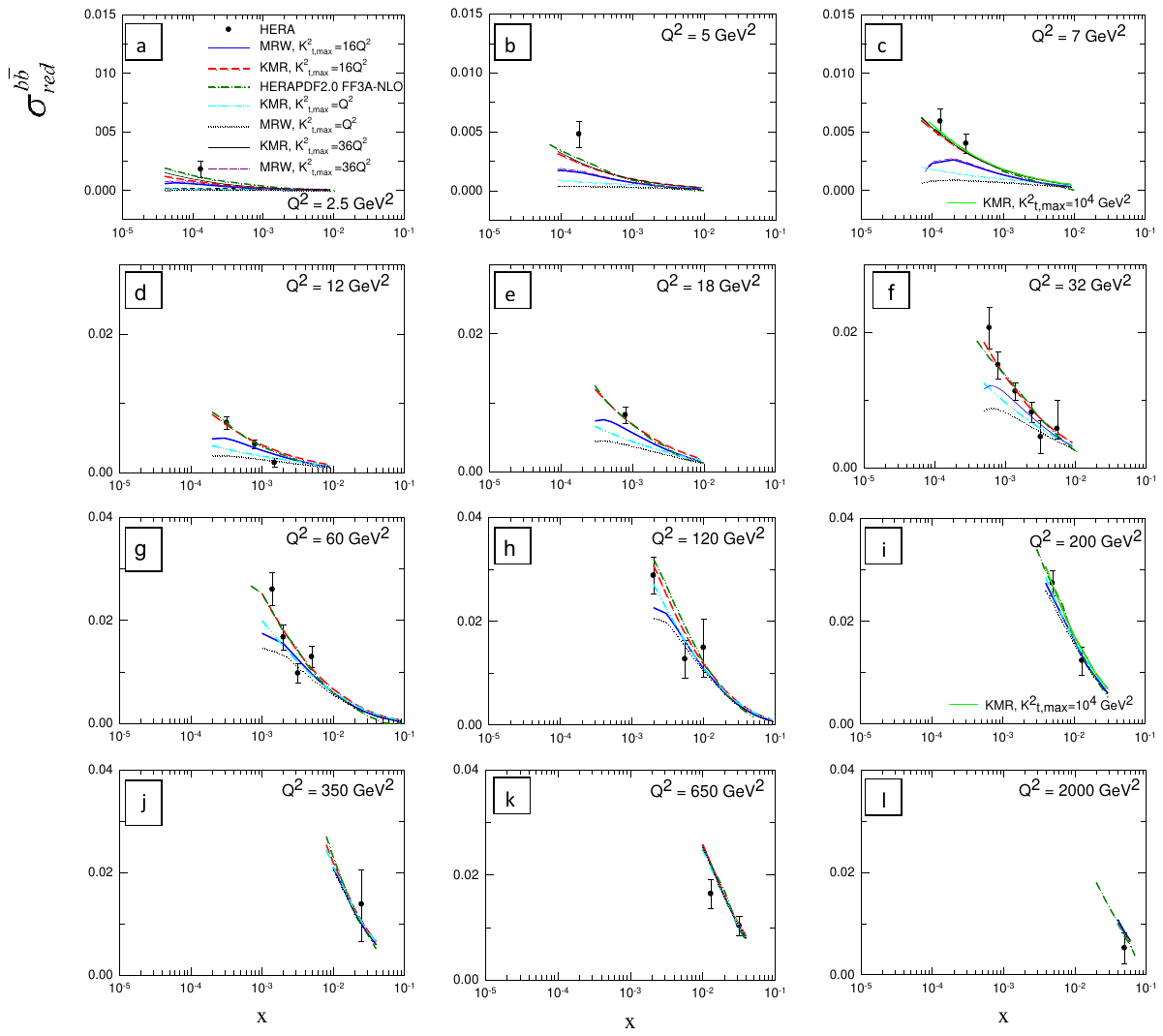}
\caption{The reduced beauty cross section as a function of $x$ for various $Q^2$ values. See the text for more explanations.}
 \label{fig:1}
\end{figure}
 
\begin{figure}[ht]
\includegraphics[width=160mm]{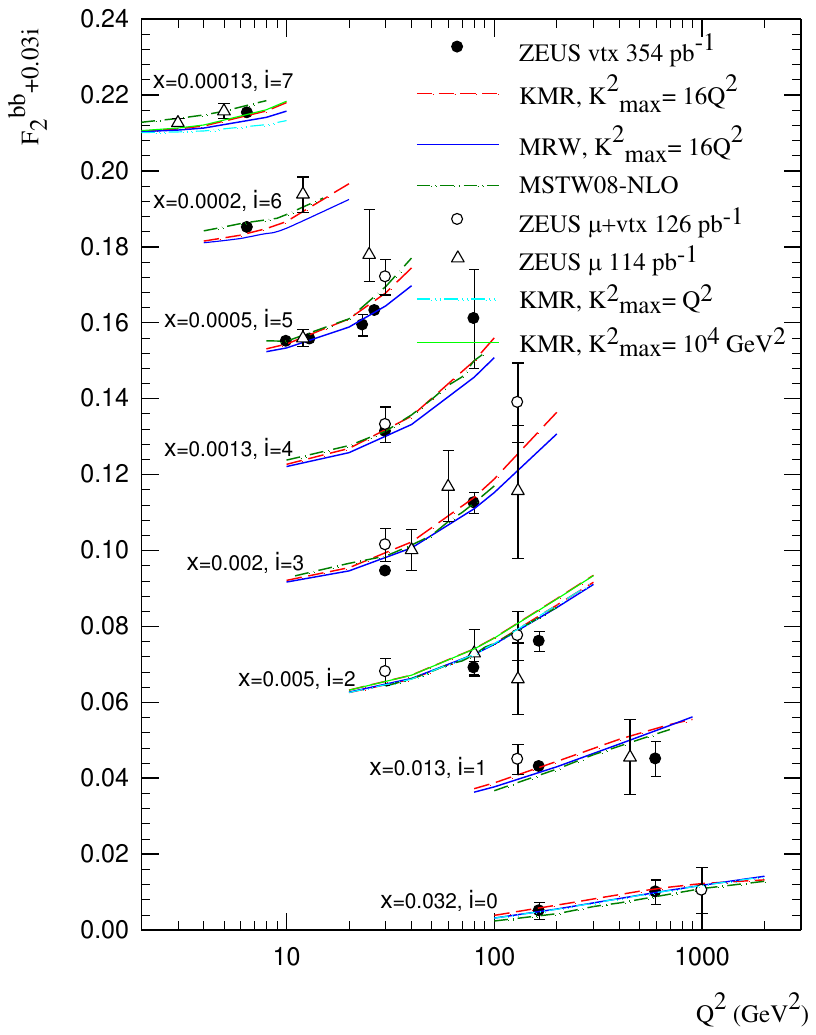}
\caption{The beauty structure function as a function of $Q^2$ for various $x$ values. See the text for more explanations.}
 \label{fig:2}
\end{figure}

\begin{figure}[ht]
\includegraphics[width=160mm]{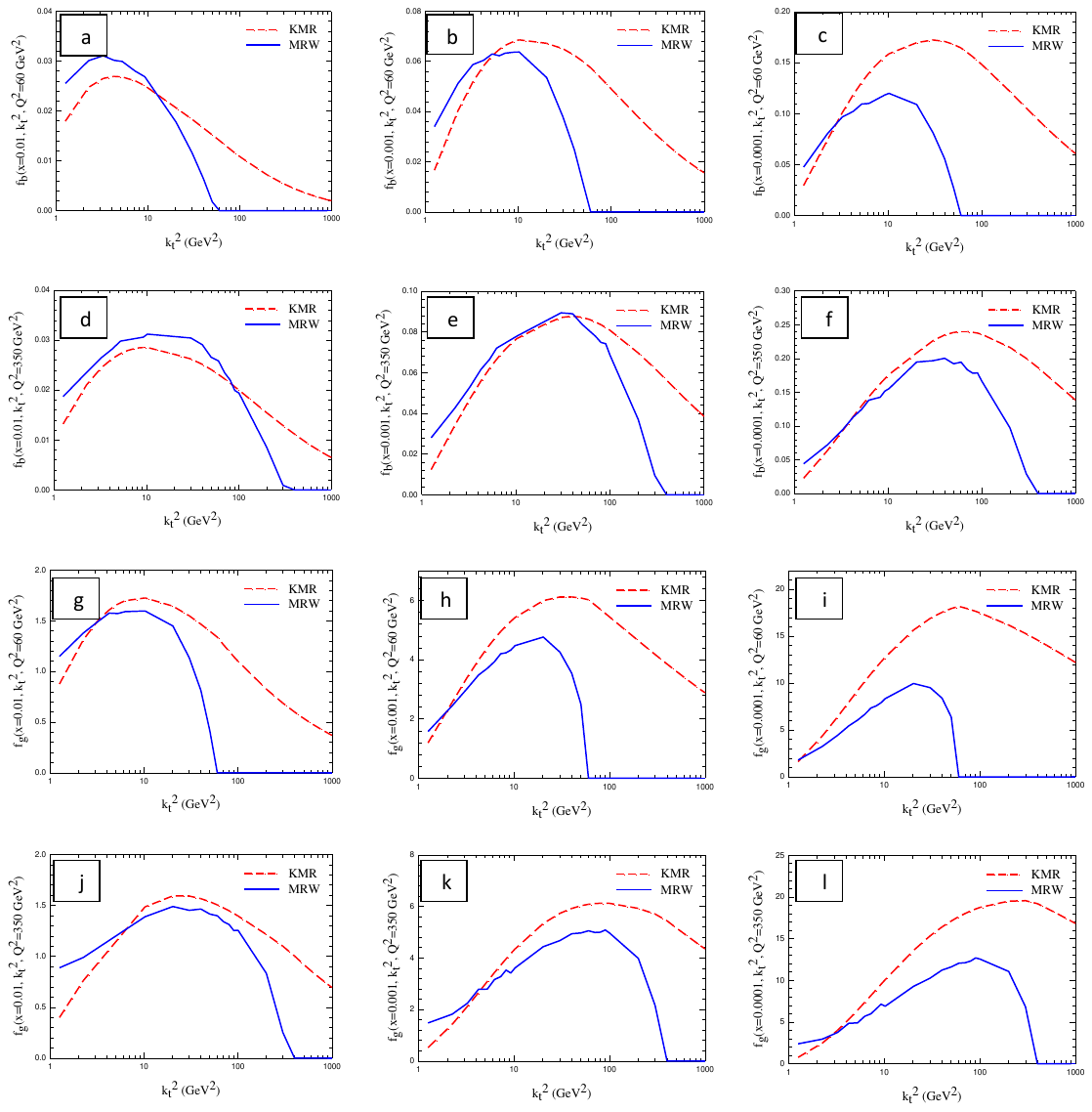}
\caption{The unintegrated beauty quark and gluon distribution functions versus ${k_t}^2$ with the $KMR$ ($MRW$) prescription by using the $MMHT2014$-$LO$ ($MMHT2014$-$NLO$) as the inputs.}
 \label{fig:4}
\end{figure}

\begin{figure}[ht]
\includegraphics[width=160mm]{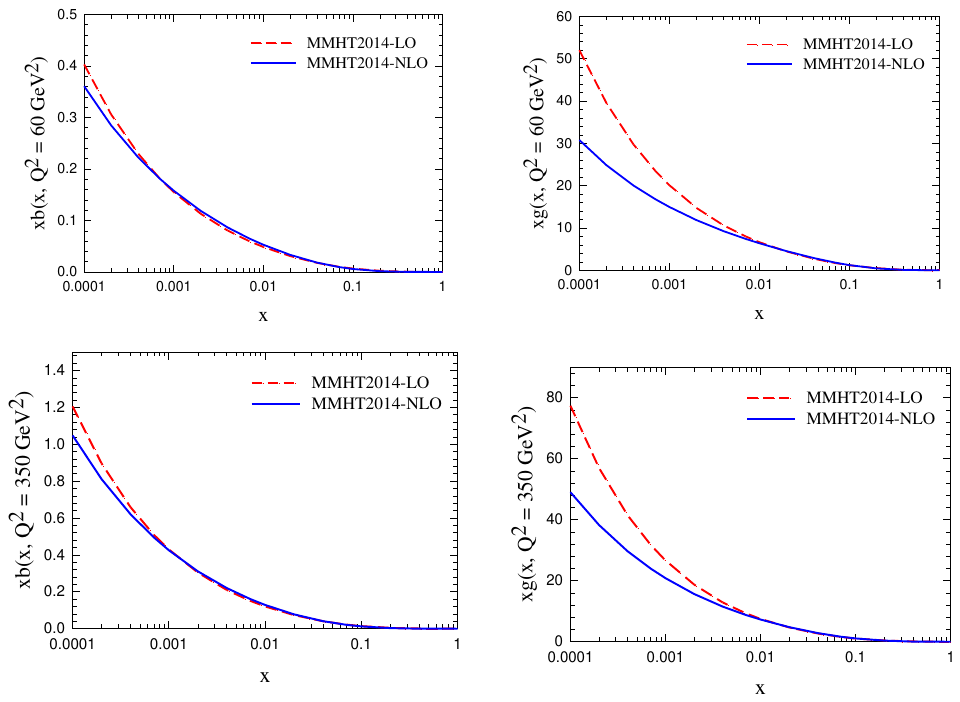}
\caption{The integrated beauty quark and gluon distribution functions at scale $Q^2=60$and $350 GeV^2$,
 by using the $MMHT2014$-$LO$ (dash curves) and $MMHT2014$-$NLO$ (full curves) \cite{22}.}
 \label{fig:5}
\end{figure}

It should be mentioned that the $k_t$-factorization is more computationally simpler than $pQCD$ and is adequate for initial investigations 
and descriptions of exclusive processes \cite{WattWZ}. The results of this paper are another confirmation of this matter. As it has been 
explained in the reference \cite{WattWZ}, we expect to reduce the discrepancy between the data and the $k_t$-factorization prediction 
by refitting the input integrated $PDF$ and using the cut-off dependent $PDF$ \cite{Lotfi} as the input for the $UPDF$.

In conclusion, the extracted
$\sigma_{red}^{b\overline{b}}(x, Q^2)$ and $F_{2}^{b\overline{b}}(x, Q^2)$ in the $k_t$-factorization formalism by
using the $KMR-UPDF$ and $MRW-UPDF$ are in a good agreement with the
predictions of the $pQCD$ and the experimental data, but those that are extracted from the $KMR$ approach,
  have a perfect agreement with the experimental data. This issue cannot be unrelated to the consideration of $T_a=1$ for $k_t>\mu$, which leads to the contribution of a $NLO$ effect in the calculation of $KMR$-$UPDF$ \cite{9}. Also, according to the study conducted on the upper limit, $k_{max}$, of the transverse-momentum integration performed in the $k_t$-factorization formalism, we hope that the computation time of the cross section at high energy region will be reduced by considering $k_{max}=Q$.
\begin{acknowledgements}
I would like to acknowledge the University of Bu-Ali Sina for their support.
\end{acknowledgements}

\end{document}